\documentclass{elsart}

\usepackage{natbib}

\usepackage{graphicx}

\usepackage{amssymb}

\newcommand{\Gcenter}[2]{
  \dimen0=\ht\strutbox%
  \advance\dimen0\dp\strutbox%
  \multiply\dimen0 by#1%
  \divide\dimen0 by2%
  \advance\dimen0 by-.5\normalbaselineskip
  \raisebox{-\dimen0}[0pt][0pt]{#2}}%

\begin{document}

\begin{frontmatter}



\title{Exploration of a 100 TeV gamma-ray northern sky using the Tibet air-shower array
combined with an underground water-Cherenkov muon-detector array}

\author[a]{T.K.~Sako\corauthref{cor}}, \ead{tsako@icrr.u-tokyo.ac.jp}
\author[a]{K.~Kawata}, 
\author[a]{M.~Ohnishi}, 
\author[b]{A.~Shiomi},
\author[a]{M.~Takita},
\author[c]{H.~Tsuchiya} 
\corauth[cor]{Corresponding author.}

\address[a]{Institute for Cosmic Ray Research, University of Tokyo, Kashiwa 277-8582, Japan}
\address[b]{College of Industrial Technology, Nihon University, Narashino 275-8576, Japan}
\address[c]{RIKEN, Wako 351-0198, Japan}

\begin{abstract}
\indent Aiming to observe cosmic gamma rays in the 10$-$1000~TeV energy region, 
we propose a 10000~m$^2$ underground water-Cherenkov muon-detector (MD) array 
that operates in conjunction with the Tibet air-shower (AS) array.
Significant improvement is expected in the sensitivity of the Tibet AS array towards celestial gamma-ray 
signals above 10~TeV by utilizing the fact that gamma-ray-induced air showers contain 
far fewer muons compared with cosmic-ray-induced ones.
We carried out detailed Monte Carlo simulations to assess the attainable sensitivity of the Tibet AS+MD array
towards celestial TeV gamma-ray signals.
Based on the simulation results, the Tibet AS+MD array will be able to reject 99.99\% of background events 
at 100~TeV, with 83\% of gamma-ray events remaining. 
The sensitivity of the Tibet AS+MD array will be $\sim20$ times better than 
that of the present Tibet AS array around 20$-$100~TeV.
The Tibet AS+MD array will measure the directions of the celestial TeV gamma-ray sources and 
the cutoffs of their energy spectra.
Furthermore, the Tibet AS+MD array, along with 
imaging atmospheric Cherenkov telescopes as well as the Fermi Gamma-ray Space Telescope and  
X-ray satellites such as Suzaku and MAXI, will make multiwavelength observations and 
conduct morphological studies on sources in the quest for evidence of the hadronic 
nature of the cosmic-ray acceleration mechanism.
\end{abstract}


\end{frontmatter}

\section{Introduction}
Supernova remnants (SNRs) are theoretically considered the most plausible candidates for acceleration 
of hadronic cosmic rays up to the knee at $\sim10^{15}$~eV in the cosmic-ray energy spectrum. 
$\pi^0$ decays following inelastic collisions between those accelerated charged cosmic rays 
and ambient matter, such as molecular clouds near the acceleration sites, are expected to produce gamma rays 
with energies as high as 100~TeV.
Meanwhile, accelerated electrons in SNRs have difficulty producing gamma rays with energies above 100~TeV
via bremsstrahlung or inverse Compton scattering, due to synchrotron cooling by magnetic field and 
the reduction of the cross section by the Klein-Nishina effect. 
Observation of the acceleration limit of cosmic gamma rays, therefore, is important to establish the 
acceleration of hadronic cosmic rays in SNRs.

The H.E.S.S. group reported on the discovery of 14 new TeV gamma-ray sources 
in the southern hemisphere in 2006~\cite{hess-apj-2006}.
Their energy spectra were measured above 200~GeV and were found to extend up to around 10~TeV, 
many of which have relatively hard spectral indices at TeV energies ranging from $-$1.8 to $-$2.8.
This implies proton acceleration in the sources, since shock acceleration models 
\cite{DSA-1, DSA-2} favor $\sim-$2 
as the indices of gamma-ray differential energy spectra at TeV energies, although
no definite conclusions have been drawn yet.
Another feature of these sources is their faintness in other wavelengths, which 
strongly stimulates unbiased celestial surveys independent of observations in other wavelengths 
using wide field-of-view apparatuses, such as air-shower arrays.

Uchiyama et al. reported on the detection of small-scale X-ray flares 
in a supernova remnant RX~J1713$-$3946, a prominent TeV gamma-ray source in the southern sky 
in 2007 \cite{uchiy-nature-2007}.
They claimed that these flares suggest an enhancement of magnetic fields in the remnant up to $\sim1$~mG,
two orders of magnitude stronger than normally assumed.
For such strong magnetic fields, the X-ray emission would be produced by a small population of electrons, 
which in turn could not account for the observed TeV flux of the remnant. Thus, 
they concluded that this provides a strong argument for acceleration of protons and nuclei to energies of 
1000~TeV and beyond. 
This statement, however, was subsequently rectified by Y. Butt et al. \cite{butt-mnras-2008},  
who contended that the speculated strong magnetic fields and the hadronic origin of the TeV gamma-ray emission 
would lead to radio synchrotron emission $\sim 1000$ times brighter than any previously observed,
if the number ratio of electrons to protons is not be unrealistically small: $< 10^{-6}$. 
They pointed out as well that even if $\sim 1$~mG magnetic fields are acceptable, 
they could not account for the TeV gamma-ray emission present over the entire SNR shell, 
insofar as they were confined within minuscule X-ray hot spots. 
Tanaka et al. \cite{tanaka-astroph-2008} claimed that the multi-wavelength spectrum of 
RX~J1713$-$3946 based on the Suzaku X-ray and the H.E.S.S. TeV gamma-ray data 
hardly can be explained within the pure leptonic model, while the hadronic model with a strong magnetic 
field provides a good fit to the spectrum, with the special arrangements of 
the model parameters being necessary to explain the lack of thermal X-ray emission. 
Drury et al. \cite{drury-astroph-2009} recently showed that the heating of the ion component of 
plasma caused by the particle acceleration in strong nonlinear shocks can be strongly suppressed,
so that thermal X-ray emission may no longer be expected.
Under these controversial circumstances, evidence for the hadronic nature of the particle acceleration 
in RX~J1713$-$3946 remains far from conclusive.

From this discussion, one can see that observations of the energy spectra of TeV gamma-ray sources 
above 10~TeV, along with morphological studies on the sources and 
positional comparison between their radio, X-ray, GeV, and TeV images,
are highly encouraged at this moment in order to obtain observational 
proof of the theoretical conjecture for hadronic cosmic-ray acceleration in SNRs.
An air-shower array with a high background cosmic-ray rejection power
that observes cosmic gamma rays in the 10$-$1000 TeV energy region at high altitude
is considered suitable for this aim. There has been, however, no such experiment so far.
In this paper, we propose a 10000~m$^2$ underground water-Cherenkov muon-detector (MD) array
that operates in conjunction with the Tibet air-shower (AS) array.

The Tibet air-shower array is located at 
Yangbajing (90.522$^\circ$~E, 30.102$^\circ$~N, 4300~m above sea level) in Tibet, China
\cite{tibet-prl-1992, tibet-apjl-1999, tibet-apj-2000}.
This array has been successfully operating, leading to a good deal of data on gamma-ray astronomy 
as well as on cosmic-ray physics: multi-TeV gamma rays from active galactic nuclei in the flaring state
such as Mrk~501 \cite{tibet-apj-2000} and Mrk~421 \cite{tibet-apj-2003}, 
the steady emission of gamma rays from the Crab Nebula \cite{tibet-apjl-1999, tibet-apj-2008a}, 
the cosmic-ray anisotropy in the sky \cite{tibet-prl-2004, tibet-apj-2005a}, 
the cosmic-ray energy spectra around the knee \cite{tibet-apj-2008b}, and so on.
In particular, it may be 
worthwhile to stress here that a few spatially separated, small-scale enhancements
were observed in the Cygnus region at multi-TeV energies \cite{tibet-sci-2006}, suggesting 
small-scale ($\sim2^\circ$) excesses of gamma rays due to some extended gamma-ray 
sources. Actually, following this result, MILAGRO reported the detection of TeV 
gamma-ray sources in the Cygnus region \cite{mgro-apjl-2007}. 

Adding the MD array to the Tibet AS array will significantly improve the sensitivity towards celestial 
gamma-ray signals above 10~TeV, by utilizing the fact that gamma-ray-induced air showers 
contain far fewer muons compared with cosmic-ray-induced ones.
The detailed structure of the MD array and the procedure of our simulation study 
to deduce its attainable sensitivity are described. 
The sensitivities of several present and future experiments are compared as well.

\section{Tibet Air-Shower Array and Muon Detector Array} \label{sec-mdarray}
\begin{figure} 
  \begin{center}
    \includegraphics[width=\textwidth]{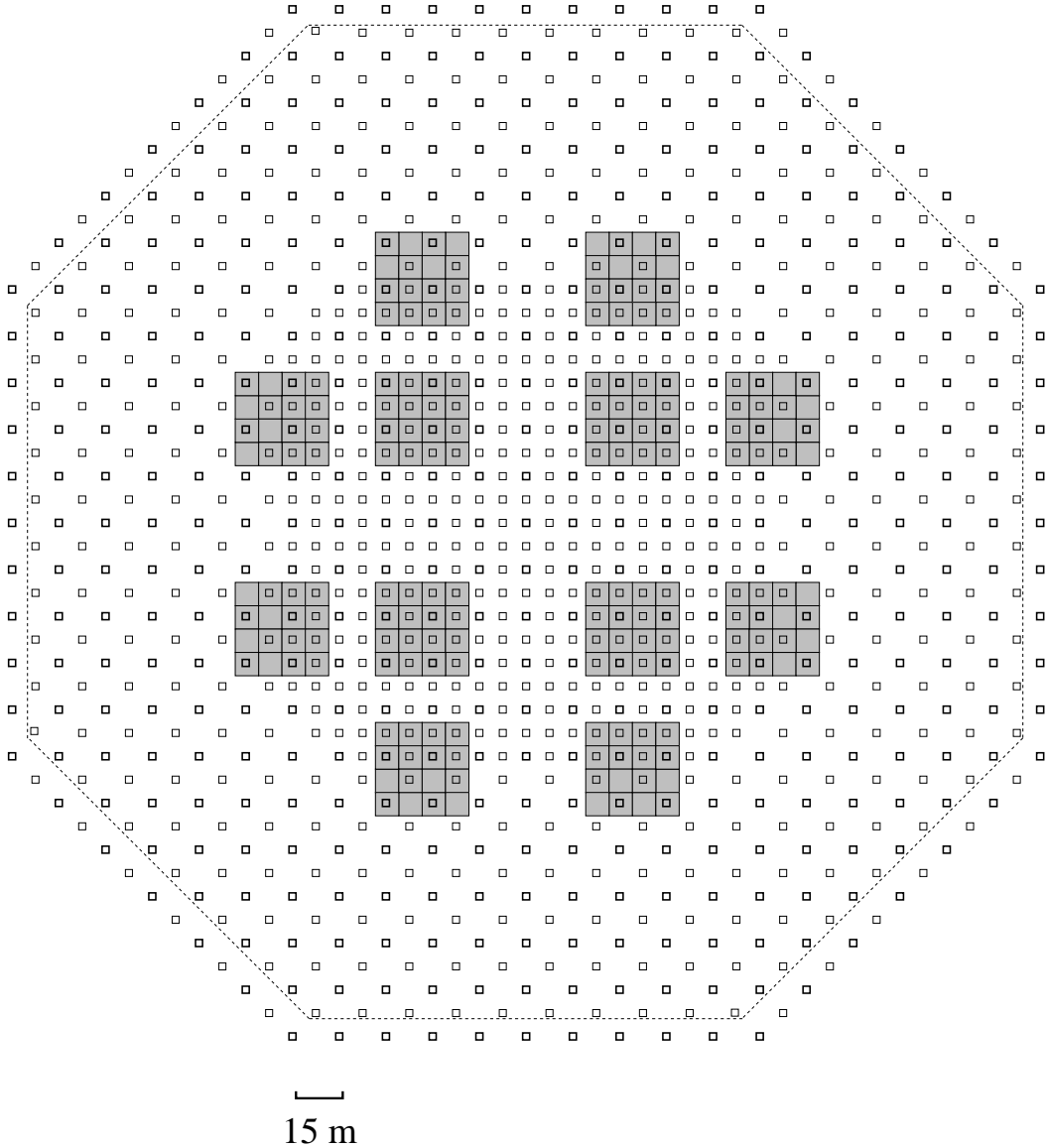} 
    \caption{Schematic view of the Tibet AS+MD array. Open squares represent surface 0.5~m$^2$
scintillation counters comprising the Tibet AS array.
Filled squares represent 52~m$^2$ pool cells of the MD array, 
set up beneath a 2.0~m thick soil and a 0.5~m thick concrete layer.
The AS array shown here is an upgraded version of the present one, consisting of 
1081 scintillation counters with an effective area of 83000~m$^2$. 
Enclosed by the dashed line is the fiducial area for selecting air shower events 
based on their core locations (See \S \ref{sec-mcsim}).}
    \label{MDarray}
  \end{center}
\end{figure}

\begin{figure}
  \begin{center}
    \includegraphics[width=\textwidth]{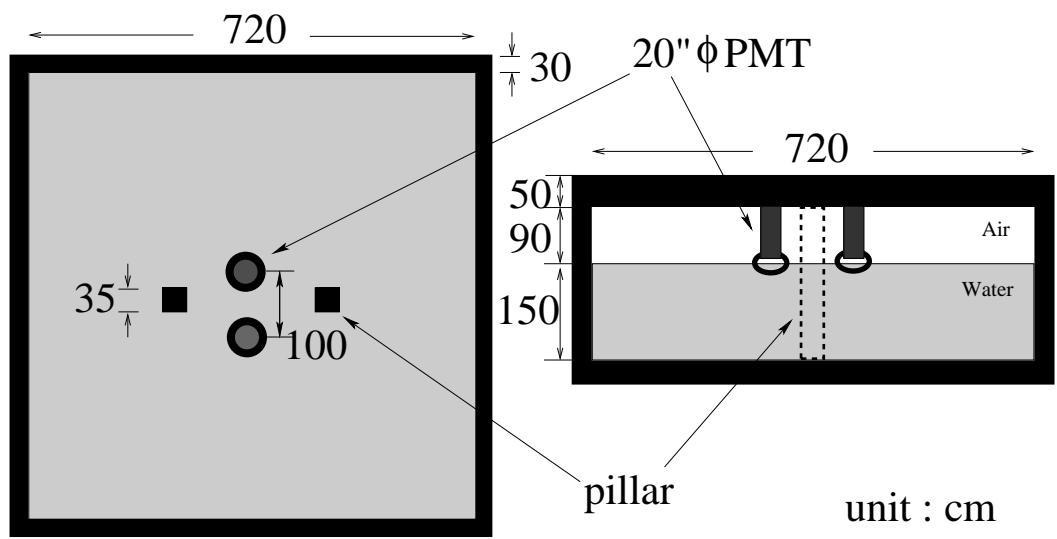} 
    \caption{Schematic view of a pool cell of the MD array (Left: top view, right: side view). 
The water volume is 7.2~m wide, 7.2~m long, and 1.5~m deep, with a 0.9~m thick air layer over it. 
Two downward-facing 20''$\phi$ PMTs are installed on the ceiling.
Two vertical pillars reinforce the structure.
}
    \label{MDcell}
  \end{center}
\end{figure}

Fig. \ref{MDarray} shows the currently proposed configuration of the Tibet AS array with
the MD array (the Tibet AS+MD array).
The surface AS array (1081 scintillation detectors, 83000~m$^2$) is a simple extension of 
the current Tibet air-shower array (789 scintillation detectors, 37000~m$^2$).
The MD array consists of 12 pools, each with 16 cells, covering a total area of 9950~m$^2$.
Each cell of the MD array, shown in Fig. \ref{MDcell}, is composed of a concrete water tank 
7.2~m wide, 7.2~m long, and 2.4~m deep, equipped with two downward-facing 20-inch-in-diameter 
(20''$\phi$) PMTs (HAMAMATSU R3600) on the ceiling.
The inner surface is coated in white paint so that it acts not only as a waterproof layer 
but also as a reflector of
Cherenkov photons, which are emitted by air-shower muons and subsequently collected by the PMTs. 
In order to suppress air-shower electromagnetic components, the MD array is 
set up beneath a 2.0~m thick layer of soil. The mass thickness of the soil and the concrete 
ceiling (0.5~m) is equivalent to $\sim$ 19 radiation lengths altogether.
The energy threshold for muons is approximately 1~GeV.

Our Monte Carlo simulations prove that the configuration of the MD cells
as described in Fig.~\ref{MDarray} provides a sensitivity enhancement of the Tibet AS array almost equal to 
that achieved by an ideal configuration, where the MD cells are uniformly distributed in the underground of 
the AS array; the sensitivity difference would be $<\sim5\%$.
Considering cost performance and construction feasibility, the Fig.~\ref{MDarray} configuration has been 
chosen.
The two vertical pillars reinforcing the cell structure (see Fig.~\ref{MDcell}) have only an insignificant
effect ($<10\%$) on the amount of the Cherenkov light detected.
Water-Cherenkov detectors significantly exceed scintillation detectors in cost performance as well as
in the ability to suppress the influence of environmental background radioactivity 
owing to the energy threshold for the production of Cherenkov photons.

The background noise of the MD array is predominated by relativistic accidental muons arriving 
at a rate of approximately 15 kHz per MD cell \cite{tibet-taup-2008}. 
This is a much higher rate than 100~Hz, the dark pulse rate of a 20''$\phi$ PMT 
measured in the laboratory at a gain of 10$^7$ under a supplied voltage of 1500~V 
and at a threshold of 10 photoelectrons.
Setting a time window of 200~ns in offline analysis 
reduces the event rate of accidental muons down to 0.6 per air shower 
(15~kHz $\times$ 16 cells $\times$ 12 pools $\times$ 200~ns).

\section{Monte Carlo simulations} \label{sec-mcsim}
Air showers are generated with the Corsika Ver.6.204 code \cite{corsika}, 
using QGSJET01c as a hadronic interaction model. 
Primary gamma rays are generated along the Crab's orbit with a differential power-law spectrum of $E^{-2.6}$ 
above 0.3~TeV.
Primary cosmic rays, the chemical composition model of which \cite{tibet-apj-2008b} is based mainly on
direct observational data, are also generated along the Crab's orbit and in the same energy range.
The number of generated air showers is $3.9\times10^7$ ($1.8\times10^8$) for gamma rays (cosmic rays).

The generated air showers are thrown at random positions within 300~m from the center of the AS array.
After simulating the response of the AS array for the air showers 
with the Epics uv8.00 code \cite{epics}, we obtain the estimated/true arrival direction, 
the core position, the sum of particle densities for all the detectors ($\Sigma\rho$), and so on
for each air shower.
The air showers that trigger the AS array 
whose cores hit more than 300~m from the center of the AS array 
are estimated to be less than 0.8\%, 0.04\%, and 0.01\% of all the air showers 
triggering the AS array at $32 \le \sum\rho < 56$, $100 \le \sum\rho < 215$, and 
$215 \le \sum\rho < 464$, respectively.
We select air showers that fulfill the following criteria: \\
(i) Any four-fold coincidence must occur in the scintillation counters recording 1.25 particles or more 
in charge. \\
(ii) Five out of the six hottest counters must be contained in the fiducial area enclosed by a dashed line 
in Fig. \ref{MDarray}. For air showers that hit fewer than 6 counters, all of the hit counters 
must be in the fiducial area. \\
(iii) The residual error in reconstructing the arrival direction by means of the least-squares method 
must be less than 1.0 m. \\
In point-source analyses, we normally extract air-shower events coming from within a search window 
around a target source.
The radius of the window is expressed as 6.9$^\circ$/$\sqrt{\Sigma \rho}$ as a function of $\Sigma\rho$, 
which maximizes the detection significance of the sources \cite{tibet-apj-2003}.
Correspondingly in this simulation analysis, we extract events such that the deviation of 
their estimated arrival direction from their true arrival direction is $< 6.9^\circ$/$\sqrt{\Sigma\rho}$. \\
The trigger efficiency for gamma rays and cosmic rays reaches 100\% above 10 and 50~TeV, respectively.
The angular and energy resolutions of air showers are estimated to be 0.2$^\circ$ and 40\%, 
respectively, for 100~TeV primary gamma rays.

\begin{figure}
  \begin{center}
    \includegraphics[width=\textwidth]{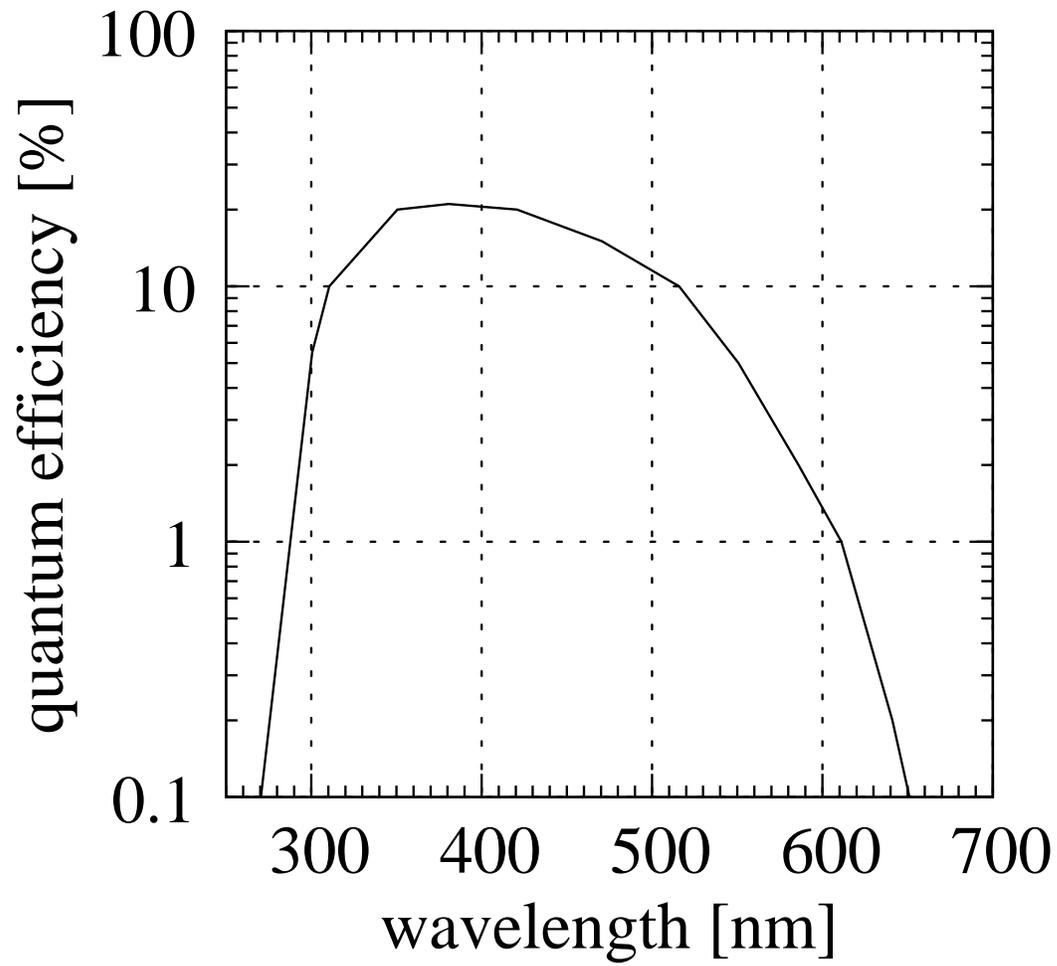} 
  \caption{Quantum efficiency of 20''$\phi$ PMTs assumed in the simulation.}
  \label{cellparam}
  \end{center}
\end{figure}

\begin{figure}
  \begin{center}
    \includegraphics[width=\textwidth]{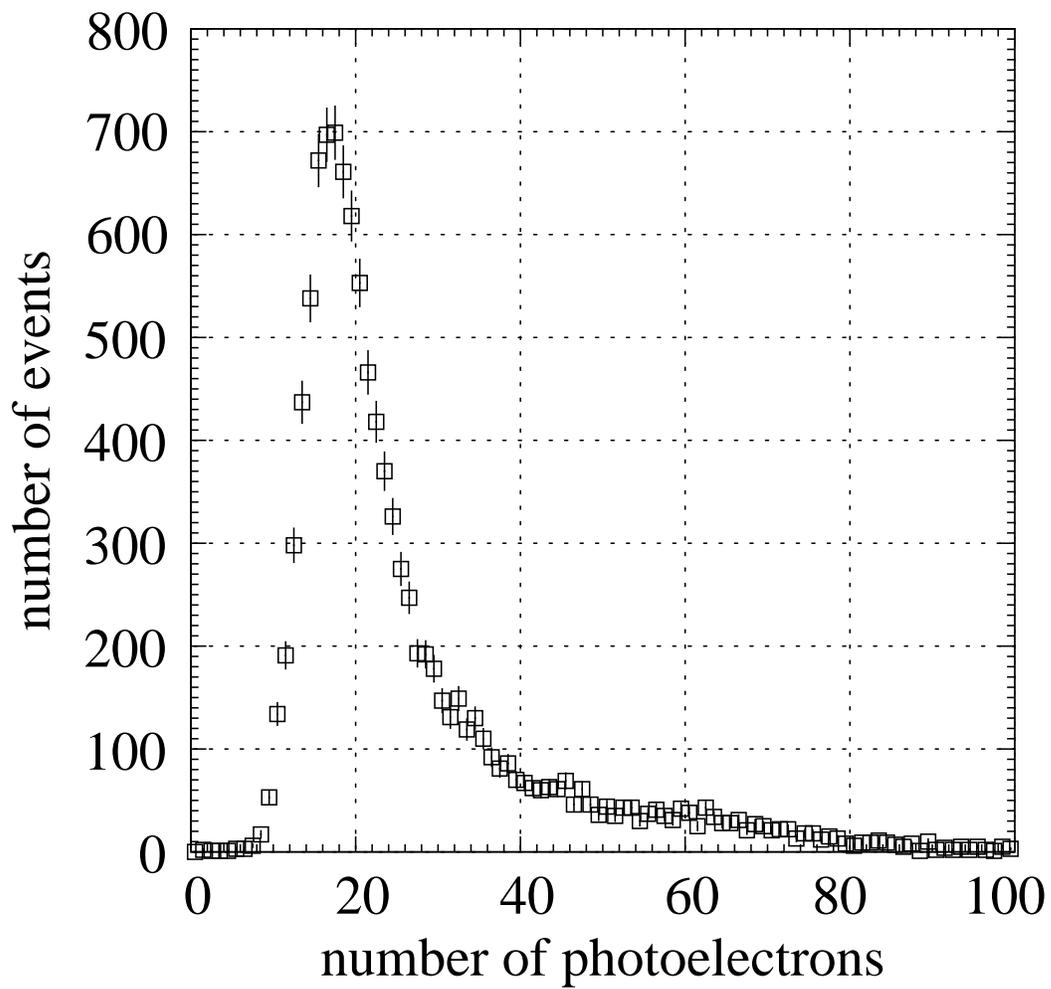} 
    \caption{Distribution of the number of photoelectrons averaged over two PMTs
($N_{\mbox{\scriptsize PE}}$) for vertically penetrating muons 
that uniformly hit the ceiling of an MD cell.}
    \label{1mu-Npe}
  \end{center}
\end{figure}

To simulate the response to the MD pool cells as well as that to the overburdening, 
which consists of 2.0~m thick soil and 0.5~m thick concrete, we adopt GEANT4~8.0 code~\cite{geant4}, 
taking into account the details of the cell structure shown in Fig. \ref{MDcell}. 
The secondary particles of the remaining air showers are traced in the soil, which is assumed to have 
a mass density of 2.0 g/cm$^3$ and is made up of 70\% SiO$_2$, 20\% Al$_2$O$_3$, and 10\% CaO. 
All the particles surviving under the soil layer with energies above their Cherenkov thresholds 
are traced further in the MD array, along with the trajectories of Cherenkov photons that they emit.
The concrete frame of the MD cells is assumed to have a density of 2.3~g/cm$^3$ and 
to be made up of the same compositions as the soil. 
The inner surface of the cells is assumed to reflect Cherenkov photons isotropically at 70\% probability.
Fig. \ref{cellparam} shows the quantum efficiency of 20''$\phi$ PMTs employed in the simulation.
The attenuation length of light in water is set to be wavelength-dependent, being 40 m at 400 nm.
In the end, the numbers of photoelectrons averaged over two PMTs in every MD cell 
($N_{\mbox{\scriptsize PE}}$) are summed up, taking into account the light collection 
efficiency of PMTs ($\sim$60\%) including the geomagnetic field effect \cite{sk-nim-1993}.
The typical value of $N_{\mbox{\scriptsize PE}}$ for vertically penetrating muons 
that uniformly hit the ceiling of an MD cell is estimated to be 17$^{+120\%}_{-12\%}$ photoelectrons, 
as shown in Fig. \ref{1mu-Npe}.
 
\section{Results and discussion}
\begin{figure}
  \begin{center}
    \includegraphics[width=\textwidth, bb=140 20 610 505]{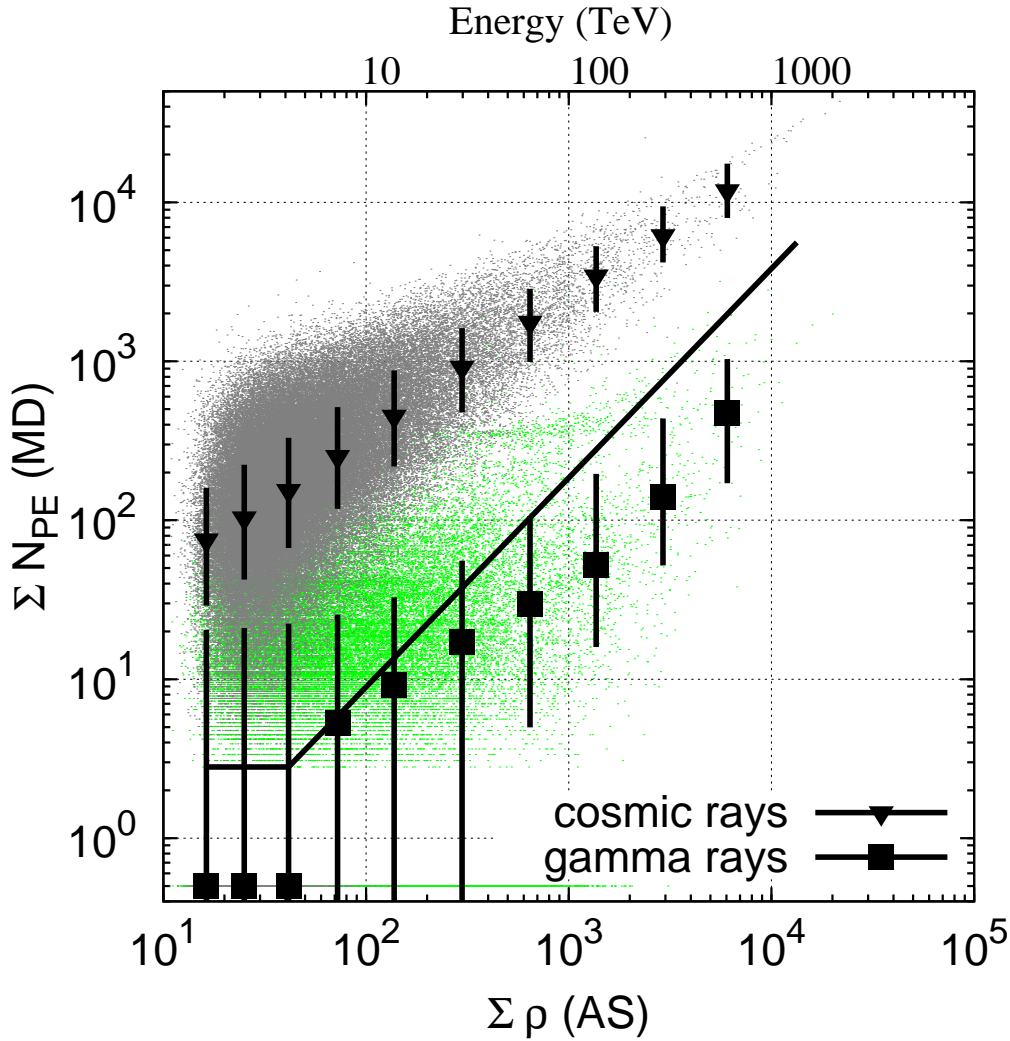} 
  \caption{Distribution of $\Sigma N_{\mbox{\scriptsize PE}}$ as a function of $\Sigma \rho$. 
$\Sigma \rho$ denotes the sum of particle densities for all the AS scintillation counters, and 
$\Sigma N_{\mbox{\scriptsize PE}}$ the sum of $N_{\mbox{\scriptsize PE}}$s from the MD pool cells.
The green and grey dots correspond to the gamma- and cosmic-ray-induced air showers, respectively. 
The filled triangles (squares) and the accompanying error bars indicate values at the median, 
20\%, and 80\% of the $\Sigma N_{\mbox{\scriptsize PE}}$ distribution of gamma-ray (cosmic-ray) events 
in each $\Sigma \rho$ bin.
The air showers producing no PEs are plotted at $\Sigma N_{\mbox{\scriptsize PE}}$ = 0.5.
The number of photoelectrons that each 20''$\phi$ PMT can detect is assumed to saturate at $\sim$300.
The number of green and grey dots plotted do not reflect the true flux ratio of the gamma rays
to the cosmic rays; for demonstration, a sufficient number of gamma-ray events are generated
so that the high-energy gamma-ray events above 100~TeV can be recognized on the figure.
Meanwhile, the fluxes of the gamma rays and the background cosmic rays under the assumption of the one-year
observation of the Crab Nebula are considered in setting the 
$\Sigma N_{ \mbox{\scriptsize PE}}$-based event-selection criterion, shown by the solid line,
which maximizes the detection significance of the source.
}
  \label{scat-plot}
  \end{center}
\end{figure}

\begin{figure}
  \begin{center}
    \includegraphics[width=\textwidth, bb=135 30 615 510]{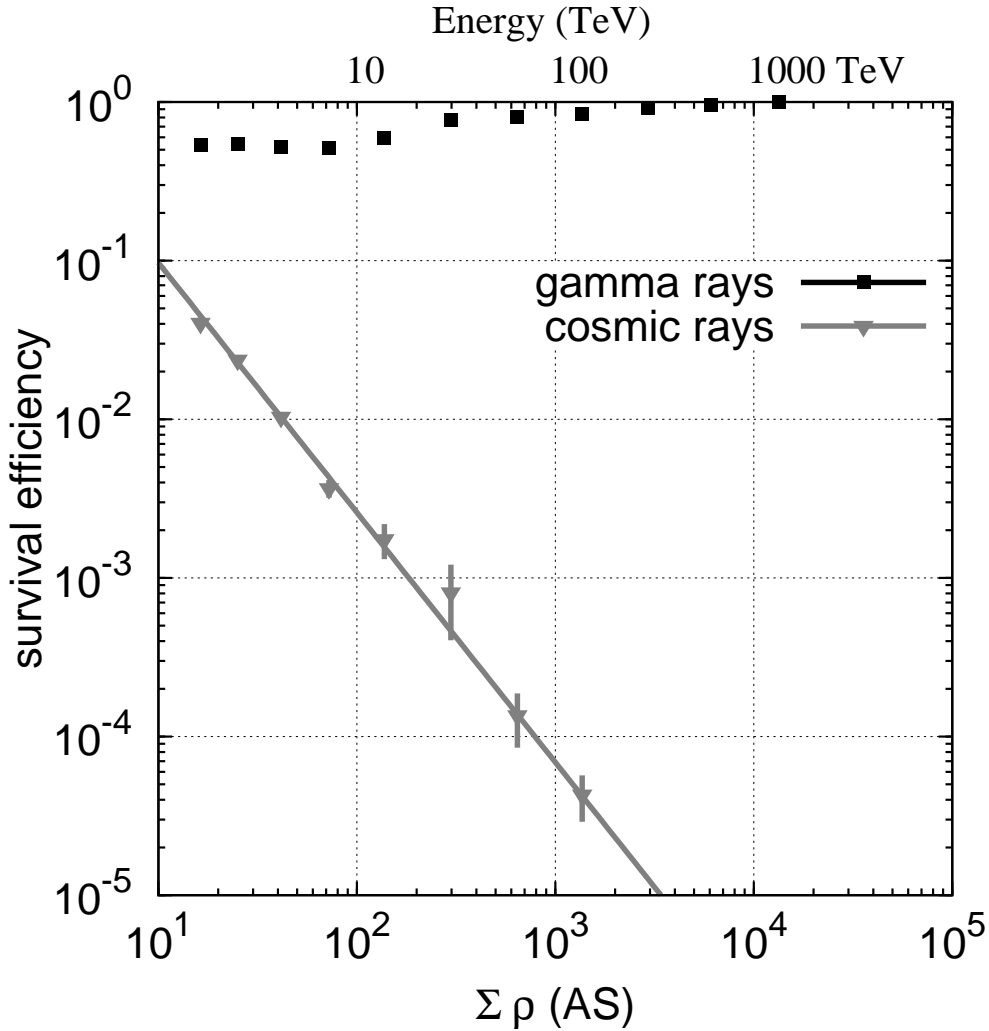}
  \caption{The survival efficiencies of gamma rays (squares) and background cosmic rays (triangles) 
after the event selection by $\Sigma N_{\mbox{\scriptsize PE}}$ as shown in Fig. \ref{scat-plot}.
Note that the survival efficiencies defined here do not include the event reduction by the search window,
which depends on the angular resolution.
The background events are rejected down to the $2\times10^{-3}$ level at 10~TeV, 
with 60\% of the gamma-ray events remaining.
The power-law fit of the cosmic-ray survival efficiencies (grey solid line) 
indicates that the background events are rejected down to the $5\times10^{-5}$ level at 100~TeV 
with 83\% of the gamma-ray events remaining.
The two points of background survival efficiency above 50~TeV were obtained from another data set of
the background MC events generated above 10~TeV.
}
  \label{suvival-eff}
  \end{center}
\end{figure}

\begin{figure}
  \begin{center}
    \includegraphics[width=\textwidth, bb=10 20 510 500]{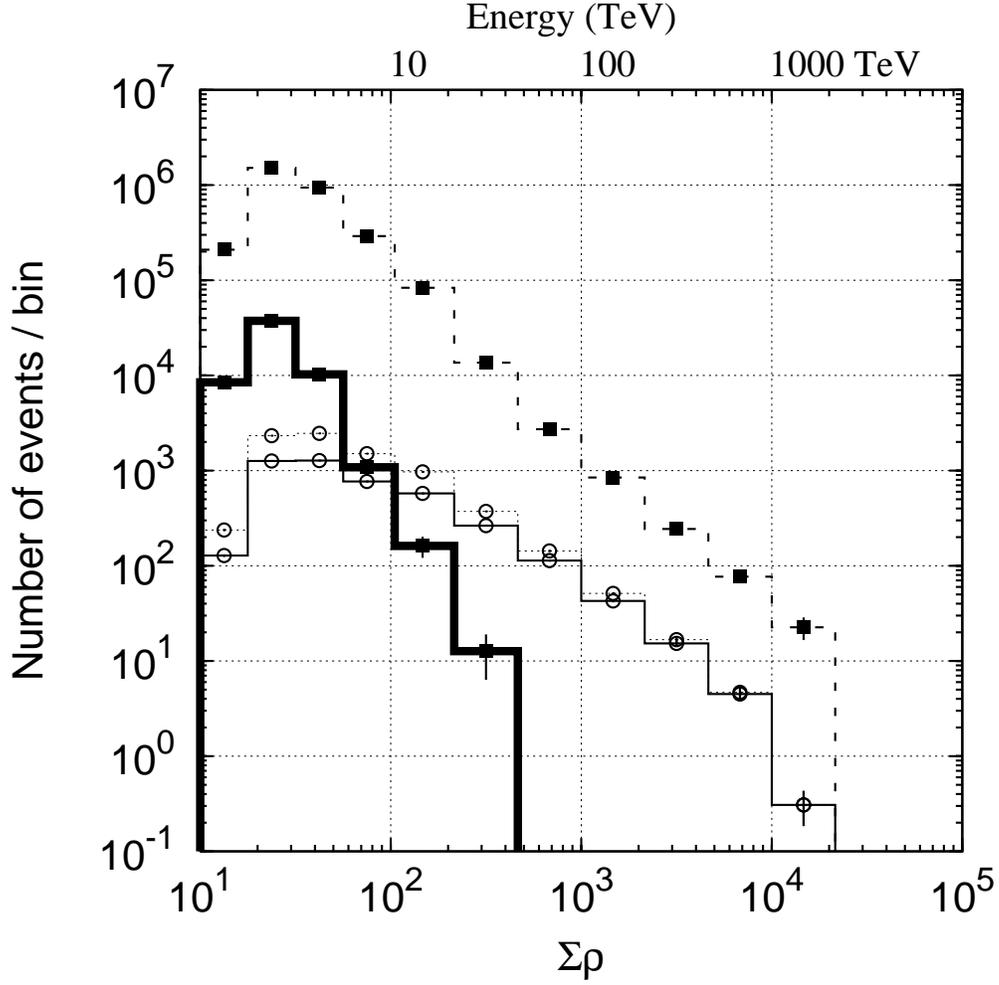}
  \caption{$\Sigma \rho$ spectrum of the gamma-ray (open circles) and background cosmic-ray (filled squares) events expected, assuming the one-year observation of the Crab Nebula with a flux of $F_{\mbox{\tiny G}} = 6.45 \times 10^{-10} \times (E/0.3\mbox{TeV})^{-2.6}$ [cm$^{-2}$s$^{-1}$TeV$^{-1}$] and the integral flux of the cosmic rays of $F_{\mbox{\tiny CR}}(>\mbox{0.3TeV}) = 1.03 \times 10^{-4}$ [cm$^{-2}$s$^{-1}$sr$^{-1}$]. The dashed (solid) lines show the spectrum before (after) the event selection by $\Sigma N_{\mbox{\scriptsize PE}}$ as shown in Fig. \ref{scat-plot}.}
  \label{Nexpdet}
  \end{center}
\end{figure}

Using the air-shower events that survive the event selection criterion described in \S\ref{sec-mcsim} 
and remain within the search window with a radius of 6.9$^\circ$/$\sqrt{\Sigma \rho}$,
we obtain the distribution of $\Sigma N_{\mbox{\scriptsize PE}}$ as a function of $\Sigma\rho$
as shown in Fig. \ref{scat-plot}.
$\Sigma N_{\mbox{\scriptsize PE}}$ denotes the sum of $N_{\mbox{\scriptsize PE}}$s from the MD pool cells
that record photoelectrons corresponding to more than 0.15 muons, whereas
$\Sigma \rho$ represents the sum of particle densities for all the AS scintillation counters. 
$\Sigma \rho = 1000$ corresponds approximately to 100~TeV for primary gamma rays.
$\Sigma N_{ \mbox{\scriptsize PE}}$ includes the noises induced by accidental muons 
(see \S \ref{sec-mdarray}) as a Poisson distribution with an average of 0.6 events per air shower.
At low energies $<~10$~TeV, more than 50\% of the gamma-ray events do not hit the MD array
($\Sigma N_{\mbox{\scriptsize PE}} = 0$), and among those with $\Sigma N_{\mbox{\scriptsize PE}} \neq 0$, 
more than 90\% are due to contamination of the accidental muons.
The $\Sigma N_{ \mbox{\scriptsize PE}}$-based event-selection criterion, 
shown by the solid line in Fig.~\ref{scat-plot}, is set so as to
maximize the detection significance of a point source, taking into account 
the fluxes of the gamma rays and the background cosmic rays under the assumption of the one-year observation
of the Crab Nebula. Selecting air-shower events under the solid line leads to the survival efficiencies 
shown in Fig. \ref{suvival-eff}. 
Note that the survival efficiencies defined here correspond to the event selection by the MD array only, 
and that they do not include the event selection by the search window, 
which depends on the angular resolution.
The background events are rejected down to the $2\times10^{-3}$ ($5 \times 10^{-5}$) level 
at 10~TeV (100~TeV), with 60\% (83\%) of the gamma-ray events remaining.
Fig. \ref{Nexpdet} shows the expected $\Sigma \rho$ spectra of the gamma-ray and background events
before and after the event selection by $\Sigma N_{\mbox{\scriptsize PE}}$ as shown in Fig. \ref{scat-plot},
assuming the one-year observation of the Crab Nebula with a flux of  
$F_{\mbox{\tiny G}} = 6.45 \times 10^{-10} \times (E/0.3\mbox{TeV})^{-2.6}$~[cm$^{-2}$s$^{-1}$TeV$^{-1}$]. 
The integral flux of the background cosmic rays is assumed to be 
$F_{\mbox{\tiny CR}}(>\mbox{0.3TeV}) = 1.03 \times 10^{-4}$~[cm$^{-2}$s$^{-1}$sr$^{-1}$].
One can see that the number of background events above $\sim$50~TeV is well suppressed down to $<$1.

\begin{figure}
  \begin{center}
    \includegraphics[width=\textwidth]{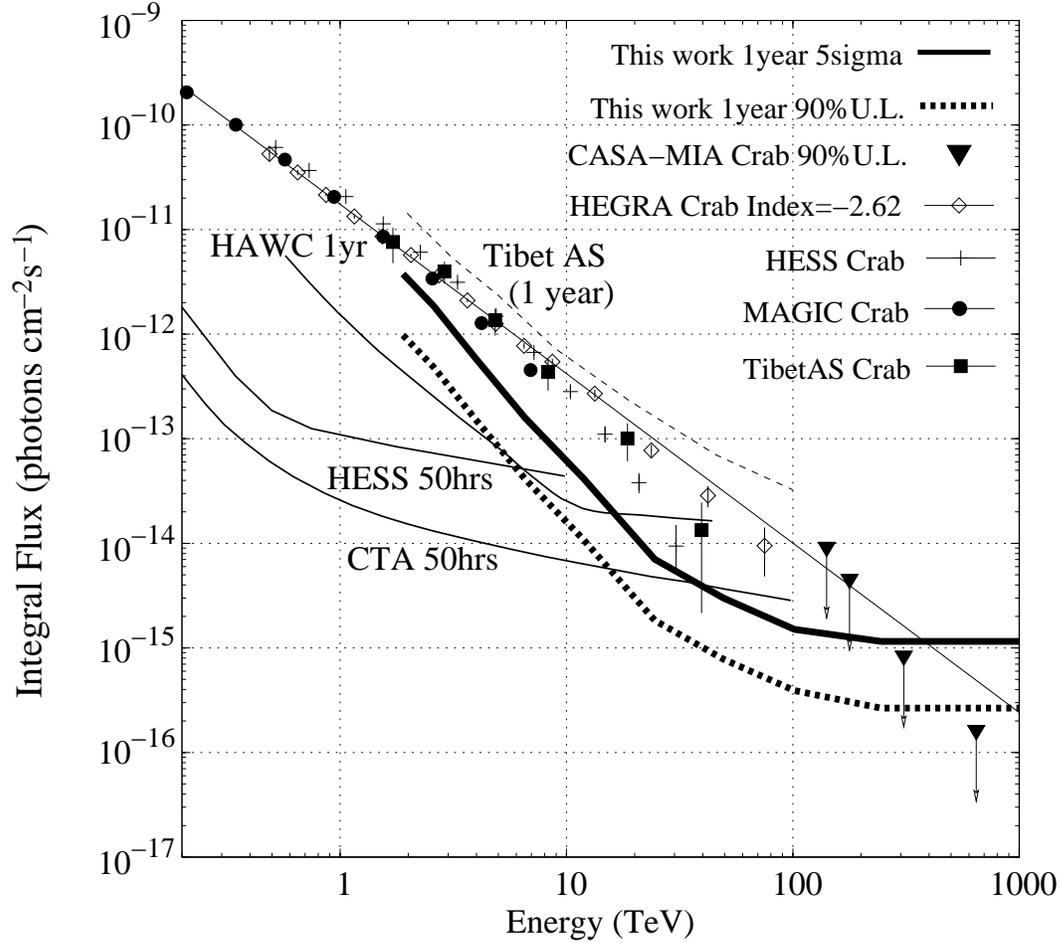} 
  \caption{The 5$\sigma$ sensitivity of the Tibet AS+MD array to a point-like gamma-ray source for
one-calendar-year observation (thick solid line), 
along with its 90\% confidence level upper limit (thick dashed line).  
Note that the sensitivities above 100~TeV, where the number of background cosmic rays is 
fully suppressed down to less than one event, is defined as a flux corresponding to 10 gamma-ray events 
per year. For reference, the sensitivity of the present Tibet AS array is drawn by the thin dashed line. 
The Crab's flux points measured by HESS \cite{hess-AA-2006}, MAGIC \cite{magic-apj-2008}, 
HEGRA \cite{hegra-apj-2004}, and Tibet AS \cite{tibet-apj-2008a} are represented by crosses, dots, 
open diamonds, and filled squares, respectively.
The triangles with downward arrows represent 90\% upper limits on the Crab's flux given 
by the CASA-MIA experiment \cite{casa-apj-1997}. 
For comparison, the 5$\sigma$ sensitivities of currently operating H.E.S.S. as well as future experiments, 
HAWC and CTA, are plotted \cite{HAWC-Sens}.}
  \label{AS-MDsens-point}
  \end{center}
\end{figure}

The integral flux sensitivity of the Tibet AS+MD array
towards a point-like gamma-ray source is calculated as shown in Fig. \ref{AS-MDsens-point}.
One can see that in one calendar year it will achieve 5$\sigma$ sensitivities
of 7\% and 20\% of the Crab flux above 20 and 100~TeV, respectively. 
Note that its sensitivity above 100~TeV, where the number of background cosmic rays is 
fully suppressed down to $\sim1$ event/(1$^\circ\times1^\circ$), 
is defined as a flux corresponding to 10 gamma-ray events. 

Table \ref{casa-comp} compares the performance of the Tibet AS+MD array with the CASA-MIA experiment
\cite{casa-apj-1997}.
The angular resolution of the Tibet AS array for primary gamma rays (0.2$^\circ$) is 10 times better 
than that of CASA-MIA at 100~TeV. Therefore, we can take a search window with a 10 times smaller radius, 
so that the solid angle of the window becomes 10$^2$ times smaller.
The effective area of the upgraded Tibet AS array (83000~m$^2$) is 1/3 that of CASA-MIA, and 
the background survival efficiency due to event selection by the MD array 
($5 \times 10^{-5}$; see Fig. \ref{suvival-eff}) will be 400 times that at 100~TeV.
Thus, we can make a rough estimation that the sensitivity of the Tibet AS+MD array will be 
$\sqrt{10^2\times(1/3)\times400} \sim 100$ times better than that of CASA-MIA at 100~TeV.
This excellent sensitivity may surpass that of existing imaging atmospheric Cherenkov telescopes 
above 12~TeV and the future Cherenkov Telescope Array (CTA) project above 40~TeV as well as 
the High Altitude Water Cherenkov (HAWC) experiment above 15~TeV.
The Tibet AS+MD array will be able to detect $\sim$10 sources such as the Crab Nebula, TeV~J2032+4130,
Mrk~421, and the MILAGRO sources \cite{mgro-apjl-2007}. \\

\begin{figure}
  \begin{center}
    \includegraphics[width=\textwidth]{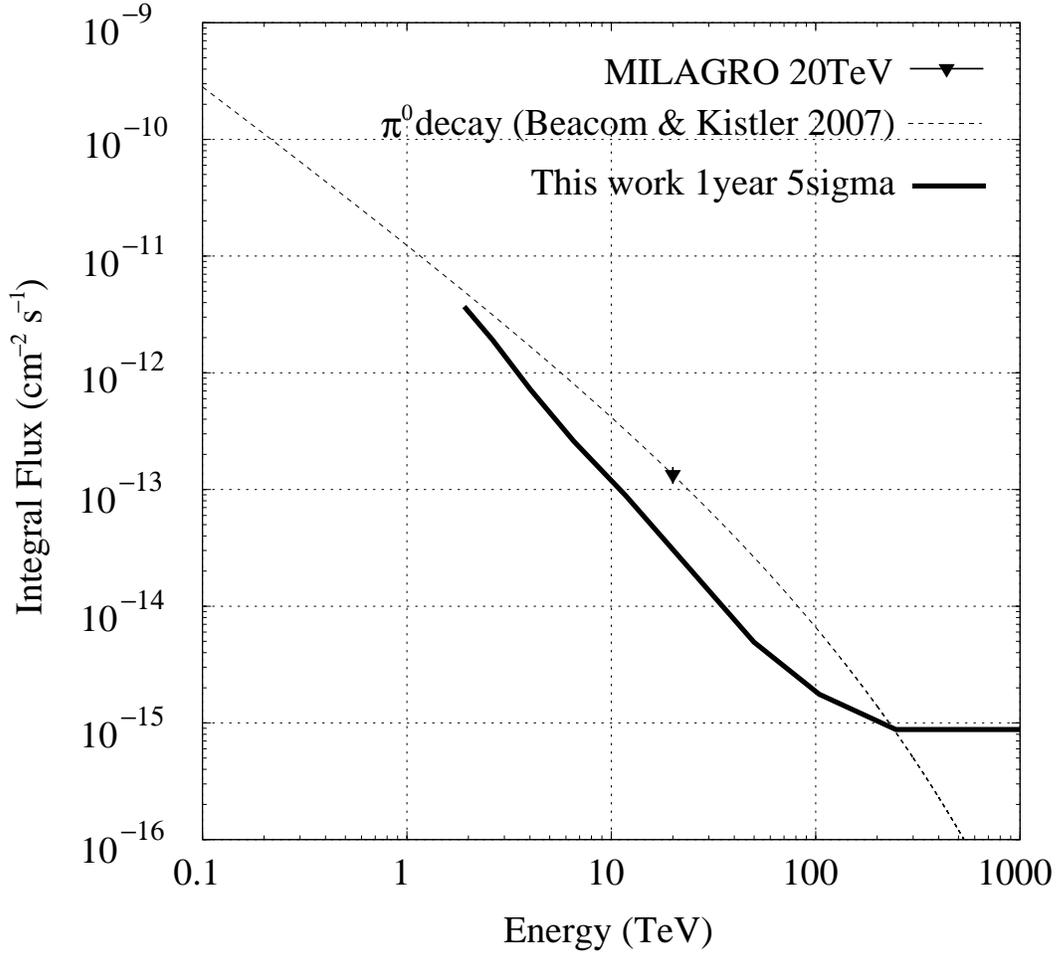} 
    \caption{The 5$\sigma$ sensitivity of the Tibet AS+MD array to a spatially extended gamma-ray source 
for one-calendar-year observation (thick solid line). 
Note that the sensitivities above $\sim$200~TeV, where the number of background cosmic rays is 
fully suppressed down to less than one event, is defined as a flux corresponding to 10 gamma-ray events
per year. The triangle represents the flux point of MGRO 2019$+$37 at 20~TeV 
measured by MILAGRO \cite{mgro-apjl-2007}. 
A $\pi^0$ decay gamma-ray flux calculated by Beacom and Kistler \cite{beacom-prd-2007} 
assuming the acceleration of protons up to 1000~TeV is shown by the dashed line.}
\label{mgro2019}
  \end{center}
\end{figure}

In order to demonstrate the sensitivity towards spatially extended sources, 
we take MGRO~J2019+37 as an example and analyze the simulation data 
using a search window with a radius of 1.7$^\circ$ independent of $\Sigma \rho$ (cf. \S \ref{sec-mcsim})
after dispersing the air-shower events randomly within a radius of 0.32$^\circ$ 
around the point source, in accordance with the MILAGRO observation \cite{mgro-apjl-2007}.
Fig. \ref{mgro2019} shows that the Tibet AS+MD array has advantages of a high signal-to-noise ratio 
and a wide field of view in observing spatially extended gamma-ray sources. 

\begin{figure}
  \begin{center}
    \includegraphics[width=\textwidth, bb=0 130 996 600]{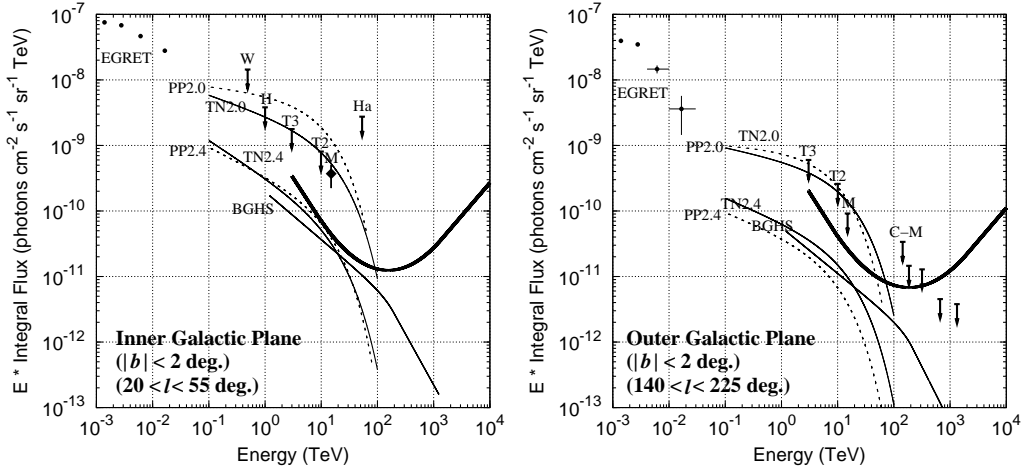} 
    \caption{
The 5$\sigma$ sensitivity of the Tibet AS+MD array towards diffuse gamma rays from the inner Galactic 
plane (left) and the outer Galactic plane (right) for a one-calendar-year observation (thick solid line).
Note that the sensitivity above $\sim$1000~TeV is defined as a flux corresponding to 10 gamma-ray events
per year. The EGRET data are represented by dots, assuming the differential 
spectral indices to be $-$2.4($-$3.3) for the inner(outer) Galactic plane \cite{EGRET-apj-1997}.
Upper limits by Whipple (W) \cite{wipple-apj-2000} with 99.9\% C.L., 
HEGRA (H) \cite{HEGRA-AA-2001} with 99\% C.L., 
HEGRA-AIROBICC (Ha) \cite{HEGRAair-app-2002} with 90\% C.L., 
CASA-MIA (C-M) \cite{CASA-apj-1998} with 90\% C.L., 
and the Tibet air shower experiment \cite{tibet-advsr-2006, tibet-astroph-2006} with 99\% C.L. 
are represented by downward arrows.
MILAGRO (M) obtained a definite flux from the inner Galactic plane and 
an upper limit with 95\% C.L. from the outer Galactic plane at 15~TeV \cite{mgro-apj-2008}. 
Theoretical curves of inverse Compton models 
by Porter and Protheroe (PP2.0 and PP2.4; dotted lines) \cite{PP-jpg-1997}, 
as well as those by Tateyama and Nishimura (TN2.0 and TN2.4; solid lines) \cite{TN-icrc-2003}, 
are included, where 2.0 and 2.4 are the source electron injection spectral indices they assumed. 
All the inverse Compton models assume an exponential cutoff of the source 
electron injection spectrum at 100~TeV.  
Theoretical curves arising from $\pi^0 \to 2\gamma$ decay
(BGHS) are given by Berezinsky et al \cite{berez-app-1993}.
}
\label{diffG}
  \end{center}
\end{figure}

Fig. \ref{diffG} shows the sensitivity of the Tibet AS+MD array towards diffuse gamma rays 
from the Galactic plane.
The Tibet AS+MD array will be able to detect the inverse Compton component from the inner Galactic plane 
within a year. Even if the inverse Compton hypotheses were wrong, 
the $\pi^0$-decay component would be detectable at $\sim$30~TeV within a few years.
The inverse Compton component from the outer Galactic plane is likely to be detectable within a year 
unless the electron injection spectral index is too soft ($\sim -2.4$). 
Positive detection of the $\pi^0$-decay component from the outer Galactic plane would necessitate
several years' observation. 
Measurements of the flux and spatial distribution of these gamma rays in the 10$-$1000~TeV energy region
will place constraints
on the model predictions, and thus make it possible to investigate the relative contribution of 
the hadronic and leptonic mechanisms.

\begin{figure}
  \begin{center}
    \includegraphics[width=\textwidth, bb=10 120 505 600]{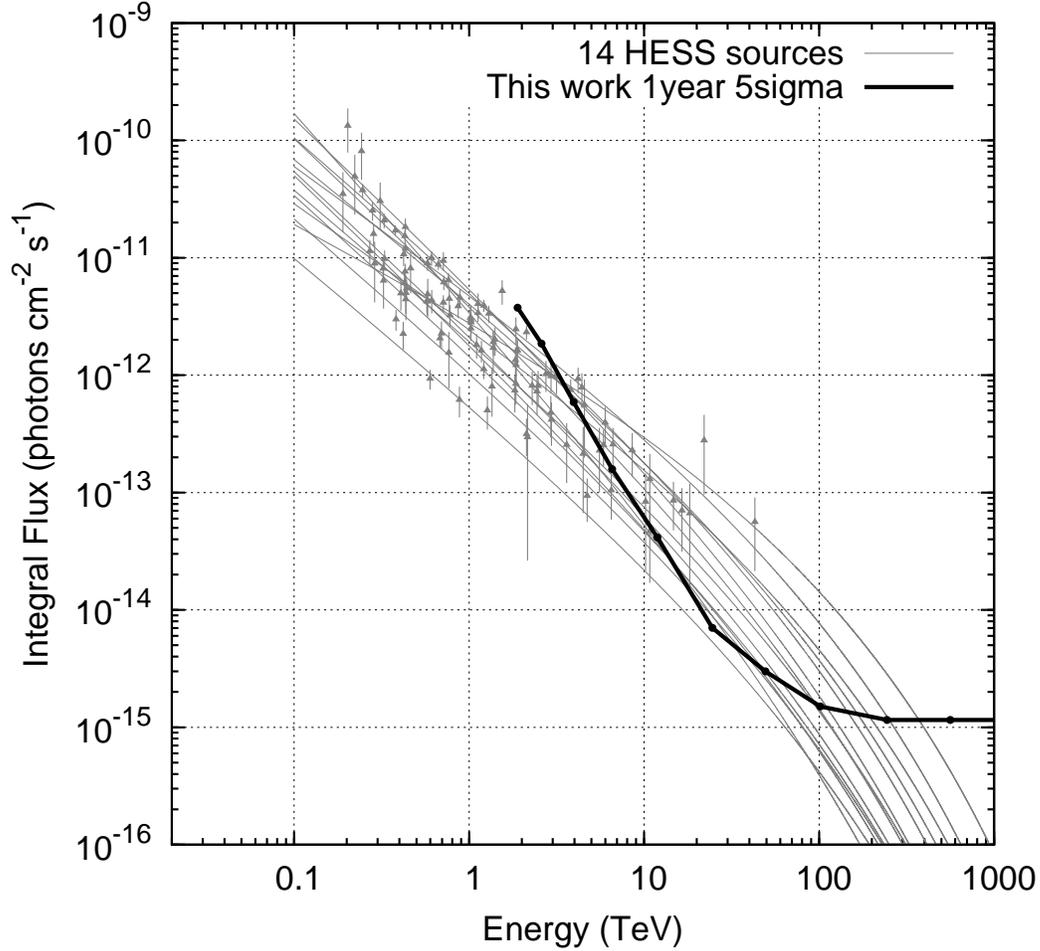} 
  \caption{The energy spectra of 14 H.E.S.S. sources reported in \cite{hess-apj-2006}, 
along with the 5 sigma sensitivity of the Tibet AS+MD array to a point-like gamma-ray source for
one-calendar-year observation (black thick line). 
The grey dots represent the integral fluxes calculated from the observed differential fluxes of the H.E.S.S.
sources point by point, assuming their spectral indices. The grey thin lines show power-law fittings to 
the H.E.S.S. data points with exponential cutoffs assuming the cutoff energy of source protons to be 
1000 TeV \cite{beacom-prd-2007}.
Note that the sensitivity above 100~TeV is defined as a flux corresponding to 10 gamma-ray events per year.}
\label{hess14}
  \end{center}
\end{figure}

Fig. \ref{hess14} shows that the Tibet AS+MD array would be able to detect 
most of the 14 H.E.S.S. sources reported in \cite{hess-apj-2006}, if located at the H.E.S.S. site. 
The H.E.S.S. experiment has discovered 46 new TeV gamma-ray sources thus far \cite{TeVcat}. 
The Tibet AS+MD array could discover some unknown TeV gamma-ray sources in the northern sky as well, 
as no extensive searches for TeV gamma-ray sources have been carried out thus far in the northern sky 
with high-sensitivity apparatuses comparable to H.E.S.S.

Lastly, multi-TeV gamma rays from intense gamma-ray bursts may be detectable, taking into account 
the absorption by intergalactic infrared photons \cite{Totani-apj-2002}, if they arise in the 
relative vicinity of the Earth ($z < \sim0.1$) and if their keV-MeV spectra extend to higher energies.

The construction of the MD array is estimated to cost roughly 5 million USD.
The Tibet AS+MD array will be a cosmic-ray experiment designed to observe 10$-$1000~TeV 
celestial gamma rays and involving the operation of the world's largest 
water-Cherenkov muon-detector array in conjunction with the surface air-shower array. 
This experiment will be complementary to H.E.S.S. and the proposed CTA project, both of which are 
located in the southern hemisphere, as well as to MAGIC, VERITAS, and the proposed HAWC experiment, 
which observe up to $\sim10$~TeV.
The Tibet AS+MD array, along with imaging atmospheric Cherenkov telescopes as well as 
the Fermi Gamma-ray Space Telescope (10~keV$-$300~GeV) and X-ray satellites 
such as Suzaku (0.4$-$600~keV) and MAXI (0.5$-$30~keV), will make multiwavelength observations and 
conduct morphological studies on sources in the quest for evidence of the hadronic 
nature of the cosmic-ray acceleration mechanism.

\section*{Acknowledgments}
We thank the members of the Tibet AS$\gamma$ collaboration for valuable discussions. 
T.K.~Sako is supported by a Grant-in-Aid for JSPS Fellows, No. 20$\cdot$5715.
For this Monte Carlo calculation we used the computers in the Institute for Cosmic Ray Research, 
the University of Tokyo.

{}

\clearpage

\begin{table}
  \caption{Performance comparison between the Tibet AS+MD array with the CASA-MIA experiment \cite{casa-apj-1997}}
  \label{casa-comp}
  \begin{center}
    \begin{tabular}{ccccccc}
      \hline
        & Energy threshold & Angular & $\mu$ det. & AS array & BG hadrons \\
        & (75\% trigger efficiency) & resolution & effective area & effective area & survival efficiency \\
\hline
Tibet   & \Gcenter{2}{$\sim$10 TeV}  & 0.2$^\circ$ & \Gcenter{2}{10000 m$^2$} & 83000 m$^2$ & \Gcenter{2}{0.005\% (100 TeV)} \\
AS$+$MD &               &  (100 TeV) &      &           (upgraded)  \\
\hline
CASA-   & \Gcenter{2}{$\sim$100 TeV} & 2$^\circ$   & \Gcenter{2}{2500 m$^2$}  & \Gcenter{2}{230000 m$^2$} & 2\% (178 TeV) \\
MIA     &               &  (100 TeV) &             &              & 0.02\% (646 TeV) \\
      \hline
    \end{tabular}
  \end{center}
\end{table}

\clearpage

\clearpage

\clearpage

\clearpage

\begin{figure}
  \begin{center}
    \includegraphics[width=\textwidth]{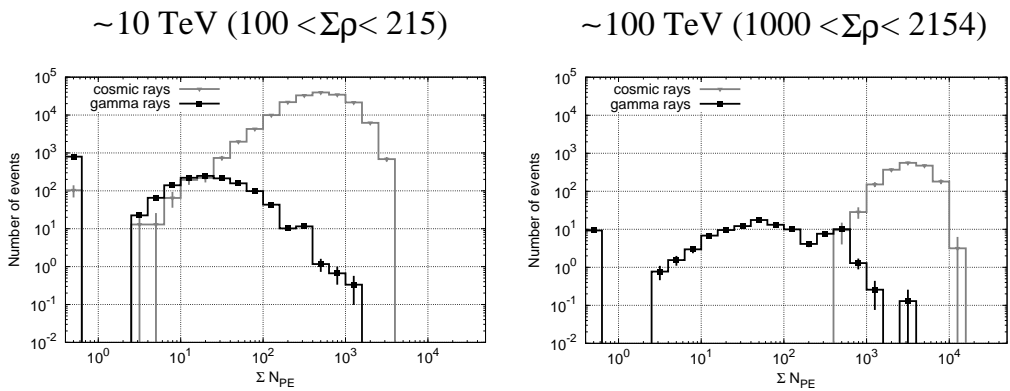} 
  \caption{
$\Sigma N_{\mbox{\scriptsize PE}}$ distribution of gamma-ray and background cosmic-ray events 
in the 10~TeV energy bin (Left; $100 \le \sum\rho < 215$) and in the 100~TeV energy bin
(Right; $1000 \le \sum\rho < 2154$), considering the fluxes of the gamma rays and 
the background cosmic rays under the assumption of the one-year observation of the Crab Nebula.
}
  \label{scat-plot-2}
  \end{center}
\end{figure}

\clearpage

\clearpage

\clearpage

\clearpage

\clearpage

\clearpage

\begin{thebibliography}{}
\bibitem{hess-apj-2006} F. Aharonian et al., Astrophys. J. 636 (2006) 777.
\bibitem{DSA-1} E. Berezhko, \& J. V\"{o}lk, Astropart. Phys. 7 (1997) 183.
\bibitem{DSA-2} E. Berezhko, \& J. V\"{o}lk, Astronomy \& Astrophysics 357 (2000) 283.
\bibitem{uchiy-nature-2007} Y. Uchiyama et al., Nature 449 (2007) 576.
\bibitem{butt-mnras-2008} Y. Butt et al., MNRAS 386 (2008) L20.
\bibitem{tanaka-astroph-2008} T. Tanaka et al., Astrophys. J. 685 (2008) 988.
\bibitem{drury-astroph-2009} L. Drury et al., Astronomy \& Astrophysics 496 (2009) 1.
\bibitem{tibet-prl-1992} M. Amenomori et al., Phys. Rev. Lett. 69 (1992) 2468.
\bibitem{tibet-apjl-1999} M. Amenomori et al., Astrophys. J 525 (1999) L93.
\bibitem{tibet-apj-2000} M. Amenomori et al., Astrophys. J 532 (2000) 302.
\bibitem{tibet-apj-2003} M. Amenomori et al., Astrophys. J 598 (2003) 242. 
\bibitem{tibet-apj-2008a} M. Amenomori et al., astro-ph/0810.3757v1, accepted by Astrophys. J.  
\bibitem{tibet-prl-2004} M. Amenomori et al., Phys. Rev. Lett. 93 (2004) 061101.
\bibitem{tibet-apj-2005a} M. Amenomori et al., Astrophys. J 626 (2005) L29. 
\bibitem{tibet-apj-2008b} M. Amenomori et al., Astrophys. J 678 (2008) 1165.
\bibitem{tibet-sci-2006} M. Amenomori et al., Science 314 (2006) 439.
\bibitem{mgro-apjl-2007} A. Abdo et al., Astrophys. J 664 (2007) L91.
\bibitem{tibet-taup-2008} M. Amenomori et al., Astrophys. Space Sci. 309 (2007) 435.
\bibitem{corsika} D. Heck et al., Report FZKA, 6019, Forschungszentrum Karlsruhe, 1998.
\bibitem{epics} K. Kasahara, http://cosmos.n.kanagawa-u.ac.jp/EPICSHome.
\bibitem{geant4} S. Agostinelli et al., Nucl. Instrum. Methods Phys. Res. A 506 (2003) 250.
\bibitem{sk-nim-1993} A. Suzuki et al., Nucl. Instrum. Methods Phys. Res. A 329 (1993) 299.
\bibitem{hess-AA-2006} F. Aharonian et al., Astronomy \& Astrophysics 457 (2006) 899.
\bibitem{magic-apj-2008} J. Albert et al., Astrophys. J 674 (2008) 1037.
\bibitem{hegra-apj-2004} F. Aharonian et al., Astrophys. J 614 (2004) 897.
\bibitem{casa-apj-1997} A. Borione et al., Astrophys. J 481 (1997) 313. 
\bibitem{HAWC-Sens} http://umdgrb.umd.edu/hawc/documents/HAWC-Sensi-3.pdf
\bibitem{beacom-prd-2007} F. Beacom \& D. Kistler, Phys. Rev. D 75 (2007) 083001.
\bibitem{EGRET-apj-1997} S. Hunter et al., Astrophys. J 481 (1997) 205.
\bibitem{wipple-apj-2000} S. LeBohec et al., Astrophys. J 539 (2000) 209.
\bibitem{HEGRA-AA-2001} F. Aharonian et al., Astronomy \& Astrophysics 375 (2001) 1008.
\bibitem{HEGRAair-app-2002} F. Aharonian et al., Astropart. Phys. 17 (2002) 459.
\bibitem{CASA-apj-1998} A. Borione et al., Astrophys. J 493 (1998) 175.
\bibitem{tibet-advsr-2006} M. Amenomori et al., Advances in Space Research 37 (2006) 1932.
\bibitem{tibet-astroph-2006} M. Amenomori et al., astro-ph/0611335v1.
\bibitem{mgro-apj-2008} A. Abdo et al., Astrophys. J 688 (2008) 1078.
\bibitem{PP-jpg-1997} T. Porter \& R. Protheroe, J. Phys. G 23 (1997) 1765.
\bibitem{TN-icrc-2003} N. Tateyama \& J. Nishimura, in: Proceedings of the 28th International Cosmic Ray Conference, Tsukuba, 4, 2003, p. 2285.
\bibitem{berez-app-1993} V. Berezinsky et al., Astropart. Phys. 1 (1993) 281.
\bibitem{TeVcat} S. Wakely \& D. Horan, http://tevcat.uchicago.edu.
\bibitem{Totani-apj-2002} T. Totani \& T. Takeuchi, Astrophys. J 570 (2002) 470.
\end{thebibliography}
\end{document}